\documentstyle[12pt]{article}
\begin{document}
\vfill \large
\title{TOWARDS THE UNIFICATION OF FUNDAMENTAL INTERACTIONS}
\author{B.G. Sidharth$^*$\\ Centre for Applicable Mathematics \& Computer Sciences\\
B.M. Birla Science Centre, Hyderabad 500 063 (India)}
\date{}
\maketitle
\footnotetext{$^*$E-mail:birlasc@hd1.vsnl.net.in}
\begin{abstract}
Recent observations of distant supernovae show that the universe, contrary
to popular belief, is expanding for ever. Similarly, recent experiments
at the Superkamiokande facility demonstrate that the Neutrino has a non
vanishing mass. Both these discoveries necessitate, at the very least, a
re-examination of conventional theories--the Big Bang Theory and the
Standard Model. On the other hand, neither has quantum gravity yielded
the desired results, nor can String Theory be taken as the last word.\\
We briefly examine this scenario, and in this light consider the
recent description of Fermions in terms of the Kerr-Newman metric and the
related model of Fluctuational Cosmology. All this is not only consistent
with known physics including the latest results alluded to, but also
explains hitherto inexplicable empirical facts like the handedness of the
Neutrino or the large number relations of Cosmology and so on. It is
shown that this scheme leads to a unified description of the fundamental
interactions. At the same time, pleasingly, we recover the usual quark
picture at the Compton wavelength scale and the Big Bang scenario at the
Planck scale.
\end{abstract}
\section{Introduction}
"....the aim is to see complete nature as different aspects of one set of
phenomena...". This is a quotation not from the Upanishads but from the
celebrated twentieth century Physicist Richard P. Feynman\cite{r1}. This has been the
goal pursued over the millennia in man's quest for an understanding of the universe. Today
looking back we can see the logic of Occam's razor or an economy of
hypothesis-- a far cry from the times of pre history when one God was
designated for each phenomenon of nature.\\
In the words of F.J. Dyson\cite{r2}, ".... the very greatest scientists in each
discipline are unifiers. This is especially true in Physics. Newton and
Einstein were supreme as unifiers. The great triumphs of Physics have been
triumphs of unification. We almost take it for granted that the road of
progress in Physics will be a wider and wider unification...".\\
However at this time when we seem to be on the threshold of the Theory of
Everything (TOE), there are a few cautionary tales which I would like to
recount, and whose significance we will encounter shortly.\\
The first is a sense of perfection and aesthetics, so much a part of
scientific philosophy from the Greeks, for example Plato to twentieth
century Physicists, for example Dirac. Thus to the Greeks the circle or
sphere were perfect, because of their all round symmetry and so all
planetary orbits had to be circular. The very word orbit is derived from
"orb" which is Greek for circle. This lead to the very complicated epicyclic model of the
Greeks which dominated science for nearly 2000 years, till almost the
17th century. As observations improved and revealed discrepancies in the
Greek model, more and more epicycles were added to bring theory in tune
with observation. It was only when Kepler around 1608 finally introduced
his elliptical orbits, that the Greek umbilical cord had finally been cut
and modern science emerged. Clearly the Greeks were trying to approximate
a single elliptical motion by a series of complicated circular motions.\\
The second incident also involves Kepler who thought that he had found the
answer to the question, "why are there only five planets besides the earth?"
Undoubtedly the creator had used the five perfect polyhedra as
celestial scaffolding, in building up the universe\cite{r3}. This idea was not just
attractive, but to Kepler it was breathtaking.\\
The third story comes from India and is recounted in the famous English poem,
"The Five Blind Men of Hindustan". Each of the blind men felt an elephant, and
ofcourse immediately understood what the elephant was like: It was like a
rope, it was like the trunk of a tree and so on.\\
Let us now come back to modern science. Sir Isaac Newton was the first great
unifier. He discovered the Universal Law of Gravitation: The force which kept
the moon going round the earth, or the earth round the sun and the force which
kept binary stars going around each other and so on were all basically the
same force of gravitation which brought apples down from a tree. This apart his Laws of Motion were also universal.\\
In the 19th Century the work of the likes of Faraday and Ampere show the close
connection between the apparently totally dissimilar forces of electricity
and magnetism. These studies culminated in the work of Maxwell who unified
not just electricity and magnetism but optics as well.\\
Another great unification of the last Century was that of Thermodynamics and the
kinetic theory of gases\cite{r4}.\\
In the early part of this Century Einstein fused space and time, giving them
an inseparable identity. He went on to unify space-time with gravitation
in his General Theory of Relativity. However he and subsequent scientists
failed in what may be called Einstein's IDEE FIXE, namely the unification of
electromagnetism and gravitation.\\
Yet another unification in this century, which often is not recognised as such is the
fusion of Quantum Mechanics and Special Relativity by Dirac, through his
celebrated equation of the electron.\\
Some thirty years ago, another unification took place due to the work of
Salam, Weinberg, Glashow and others-- the unification of electromagnetism
with the weak forces. This is the story to which we now come.\\
The weak force is one of two forces, the other being the strong force,
discovered during this Century itself. Studies and work earlier this Century
revealed that there seem to be three basic particles in the universe, the
protons, the neutrons and the electrons.\\
While the proton and the electron interact via the electromagnetic force, in
the absence of this force the proton and the neutron appear to be a pair or
a doublet, because now the distinction between the electrically charged
proton and the neutral neutron disappears. However the proton and the neutral neutron interact
via "strong forces", forces which are stronger than the electromagnetic but
have a much shorter range of just about $10^{-13}cms$.\\
The existence of the neutrino was postulated by Pauli in 1930 to explain the
decay of the neutron, and it was discovered by Reines and Cowan in 1955.
The weak force is associated with neutrino type particles and has an even
shorter range, $10^{-16}cms$. The neutrino itself has turned out to be one of
the most enigmatic of particles, with peculiar characteristics, the most
important of which is its handedness. This handedness property appears to
be crucial for weak forces.\\
Later work revealed that while particles like the electron and neutrino, namely
the leptons may be truly elementary, other particles like protons
and neutrons may be composite, infact made up of
still smaller objects called quarks -- six in all\cite{r5}. Today it is believed that
the quarks interact via the strong forces.\\
It must be mentioned that all these "particles", or loosely material particles
are fermions, that is they have half integral spin. It was realized that the
forces or interactions on the other hand are mediated by messengers like
photons which are bosons, that is, have integral spin.\\
According to contemporary thinking, the unification of the various interactions
relies on the fact that the messengers have spin one, that is are gauge forces,
where as in this sense gravitation is not a gauge force as it is thought to be
mediated by spin 2 particles.\\
To picturize the above let us consider the interaction between a proton and
an electron.

\vskip 10em

\noindent A proton could be imagined to emit a photon which is then absorbed by the
electron. \\
Instead of a single mediating particle we could think of multiplets, all
having equal masses. With group theoretical inputs, one could shortlist,
singlets with one particle like the photon triplets,
octets and so on as possible candidates.\\
Motivated by the analogy of electromagnetism mediated by the spin one photon,
it was realized in the 1950s that the $W^+$ and $W^-$ bosons could be possible
candidates for the mediation of the weak force. However there had to be one
more messenger so that there would be an allowable triplet alluded to earlier.
It was suggested by Ward and Salam that the third candidate could be the
photon itself, which would then provide not only a description of the weak
force but would also unify it with electromagnetism. However while the
$W$ particles were massive, the photon was massless so that they could not
form a triplet. So a heavy photon or $Z^0$ was postulated to make up a
triplet, while the photon was also used for the purpose of unification, and
moreover a mixing of $Z^0$ and the photon was required for what has been
called renormalization, that is the removal of infinities.\\
The question was how could the photon be massless while the $W$ and $Z$
particles would be massive? It was suggested that this could be achieved
through the spontaneous breaking of symmetry. For example a bar magnet when
heated, looses its magnetism, that is it becomes symmetric because one
cannot distinguish between the North and South poles. However as the magnet
cools down, the magnetism or the polarity or assymetry is regained spontaneously.
In other words there would be a phase transition (from symmetry to asymmetry).\\
In our case, before the spontaneous breaking of symmetry or the phase transition,
the $W$s, $Z$s and the photon would all be massless. After the phase transition,
while the photons remain mass less, the others would acquire mass. This phase
transition would occur at temperatures $\sim 10^{15^\circ}$ Centigrade. At
higher temperatures there would be a single electroweak force. As the
temperature falls to the above level electromagnetism and weak forces would
separate out.
\noindent Clearly the direction to proceed appeared to be to identify the guage character
of the strong force-- mediated by spin one particles, the gluons. This force
binds the different quarks to produce the different elementary particles,
other than the leptons. This is the standard model. However we have not
yet conclusively achieved a unification of the electroweak force and the
strong force\cite{r6}.\\
One proceeds by analogy with the electroweak unification to
obtain a new guage force that has been called by Jogesh Pati and Abdus
Salam as the electro nuclear force, or in a similar scheme the Grand
Unified Force by Glashow and Georgi. It must be mentioned that one of the
predictions is that of the proton would decay with a long life time of about $10^{32}$ years,
very much more than the age of the universe itself. However we are near
a situation where this should be observable.\\
The recent super Kamiokande determination of neutrino mass is the first
evidence of Physics beyond the standard model. Interestingly in this theory
we would also require a right handed neutrino in this case.\\
Meanwhile extended particles had come into vogue from about twenty years with
string theory. Starting off with objects of the size of the Compton
wavelength, the theory of superstrings now deals with the Planck length of
about $10^{-33}cms$.  We have noted that all interactions except gravitation are generalizations
of the electromagnetic guage theory. String theory combines
Special Relativity, Quantum Mechanics and General Relativity - we need ten,
$(9+1)$, dimensions for quantizing strings, and we also get a mass less particle of
spin two which is the mediator of the gravitational force. This way there is
the possibility of unifying all interactions including gravitation. Further,
in the above ten dimensions there are no divergences-- this is because the spatial
extension of the string fudges the singularities. However, we require, for
verification of the string model, energies $\sim 10^{18} m_P,$ as against the
presently available $10^3m_P$.\\
Another interesting feature of string theory is \underline{duality}. There are five different
solutions (compactifications) leading to the same physical picture
reminiscent of the "Five Blind Men of Hindustan". These five theories are but
different descriptions of a single theory.\\
The exotic dimensions, the large number of solutions and the non verifiable
nature of the theory are some of the unsatisfactory features of this
development, just as the 18 arbitrary parameters\cite{r7}, and the unseen monopoles
are some of the unsatisfactory features of the standard model.
\section{A New Model Towards Unification}
In previous communications\cite{r8,r9,r10} a model for leptons as Kerr-Newman type
Black Holes was developed, and it was pointed out that several hitherto
inexplicable features turned out to have a natural explanation, for
example the quantum of charge, the electromagnetic-gravitational interaction
ratio, the handedness of the neutrino and so on. We briefly recapitulate some
relevant facts.\\
It must be mentioned that the possibility that a
particle could be a Schwarzchild black hole has been examined earlier by
Markov, Motz and others\cite{r11,r12} and leads to a high particle mass of
$10^{-5}gm$, the Planck mass without much insight into other properties.\\
So let us
approach the problem from a different angle. We consider a charged Dirac
(spin half) particle. If we treat this as a spinning black hole, there is an
immediate problem:The horizon of the Kerr-Newman black hole becomes in
this case, complex\cite{r9},
\begin{equation}
r_+ = \frac{GM}{c^2} + \imath b_,b \equiv (\frac{G^2 Q^2}{c^8} + a^2 -
\frac{G^2M^2}{c^4})^{1/2}\label{e1}
\end{equation}
where $G$ is the gravitational constant, $M$ the mass and $a \equiv
L/Mc,L$ being the angular momentum. That is, we have a naked singularity
apparently contradicting the cosmic censorship conjecture. However, in the
Quantum Mechanical domain, (\ref{e1}) can be seen to be meaningful.\\
Infact, the position coordinate for a Dirac particle is
given by Dirac\cite{r13}
\begin{equation}
x = (c^2p_1 H^{-1}t + a_1) + \frac{\imath}{2}c\hbar
(\alpha_1 - cp_1 H^{-1})H^{-1},\label{e2}
\end{equation}
and similar equations for $y$ and $z$ where $a_1$ is an arbitrary constant and $c\alpha_1$ is the velocity
operator with eigen values $\pm c$. The real part in (\ref{e2}) is the
usual position while the imaginary part arises from zitterbewegung.
Interestingly, in both (\ref{e1}) and (\ref{e2}), the imaginary part is
of the order of $\frac{\hbar}{mc}$, the Compton wavelength, and leads to
an immediate identification of these two equations. We must remember that
our physical measurements are gross as indeed is required by the Uncertainity
Principle - they are really
measurements averaged over a width of the order $\frac{\hbar}{mc}$.
Similarly, time measurements are imprecise to the tune $\sim \frac{\hbar}
{mc^2}$. Very precise measurements if possible, would imply that all
Dirac particles would have the velocity of light, or in the Quantum
Field Theory atleast of fermions, would lead to divergences. (This is
closely related to the non-Hermiticity of position operators in
relativistic theory as can be seen from equation (\ref{e2}) itself.
Physics as pointed out begins after an averaging over the above
unphysical space-time intervals. In the process  (cf.ref.
\cite{r9}), the imaginary or non-Hermitian part of the position operator
in (\ref{e2}) disappears. That is
the above naked singularity is fudged or shielded by a Quantum Mechanical
censor, namely the minimum Compton scales.\\
We observe that if we adhoc treat a Dirac particle as a Kerr-Newman
black hole of mass $m$, charge $e$ and spin $\frac{\hbar}{2}$. The
gravitational and electromagnetic fields at a distance are given by
(cf.ref.\cite{r14}),
\begin{eqnarray}
\Phi (r) = - \frac{Gm}{r} + 0 (\frac{1}{r^3}) E_{\hat r} = \frac{e}{r^2} +
0 (\frac{1}{r^3}),E_{\hat \theta} =
0 (\frac{1}{r^4}),E_{\hat \phi} = 0,\nonumber \\
B_{\hat r} = \frac{2ea}{r^3} cos \theta + 0 (\frac{1}{r^4}),
B_{\hat \theta} = \frac{easin\theta}{r^3} + 0(\frac{1}{r^4}),
B_{\hat \phi} = 0,\label{e3}
\end{eqnarray}
which is correct. Infact, (\ref{e3}) also leads to the
electron's anomalous gyromagnetic ratio $g = 2.$!\\
We next examine more closely, this identification of a Dirac particle with
a Kerr-Newman black hole. We reverse the arguments after equation (\ref{e2})
which lead from the complex or non-Hermitian coordinate operators to Hermitian
ones: We consider instead the displacement,
\begin{equation}
x^\mu \to x^\mu + \imath a^\mu\label{e4}
\end{equation}
and first consider the temporal part, $t \to t + \imath a^0$, where
$a^0 \approx \frac{\hbar}{2mc^2}$, as before. That is, we probe into the
QMBH or the zitterbewegung region inside the Compton wavelength as suggested
by (\ref{e1}) and (\ref{e2}). Remembering that $|a^\mu | < < 1$, we have,
for the wave function,
$$\psi (t) \to \psi (t+\imath a^0) = \frac{a^0}{\hbar} [\imath \hbar
\frac{\partial}{\partial t} + \frac{\hbar}{a^0}]\psi (t)$$
As $\imath \hbar \frac{\partial}{\partial t} \equiv p^0$, the usual fourth
component of the energy momentum operator, we identify, by comparison with
the well known electromagnetism-momentum coupling, $p^0 - e\phi$, the
usual electrostatic charge as,
\begin{equation}
\Phi e = \frac{\hbar}{a^0} = mc^2\label{e5}
\end{equation}
In the case of the electon, we can verify that the equality (\ref{e5})
is satisfied:\\
We follow the classical picture of a particle as a rotating shell with
velocity $c$, as indeed is suggested by the Compton wavelength cut off
and which will be
further justified in the sequel. The electrostatic potential inside a
spherical shell of radius $'a'$ is,
\begin{equation}
\Phi = \frac{e}{a}\label{e6}
\end{equation}
As is well known, the balance of the centrifugal and Coulomb forces gives,
for an electron orbiting another at the distance $a$,
$$a = \frac{e^2}{mc^2},$$
which is the classical electron radius.\\
So, (\ref{e6}) now gives,\\
$$e\Phi = mc^2,$$
which is (\ref{e5}).\\
If we now use the usual value of $'a'$ viz., $2.8 \times 10^{-13} cm.,$
while rewriting $mc^2$ as $\hbar c/(\hbar/mc)$ and substitute the value of the electron Compton
wavelength, $\frac{\hbar}{mc} = 3.8 \times 10^{-11} cm.,$ we gets from the
above
$$\hbar c \approx 136 e^2$$
That is, we get the rationale for this fundamental relation, which no longer
turns out to be accidental. In any case, we can now see the
connection between the charge, mass and the velocity of light.\\
(It may be noted in passing that in the usual displacement operator theory
(\cite{r13}) the operators like $\frac{d}{dx}$ or $\frac{d}{dt}$
are indeterminate to the extent of a purely imaginary additive constant
which is adjusted against the Hermiticity of the operators concerned).\\
We next consider the spatial part of (\ref{e4}), viz.,
$$
\vec x \to \vec x + \imath \vec a, \mbox{where} |\vec a | = \frac{\hbar}
{2mc},$$
given the fact that the particle is now seen to have the charge $e$ (and mass
$m$). As is well known\cite{r15}, this leads in General Relativity
from the static Kerr metric to the Kerr-Newman metric where the gravitational
and electromagnetic field of the particle is given by (\ref{e3}), including
the anomalous factor $g = 2$. In General Relativity, the complex transformation
(\ref{e4}) and the subsequent emergence of the Kerr-Newman metric has no
clear explanation. Nor the fact that, as noted by Newman\cite{r16} spin is
the orbital angular momentum with an imaginary shift of origin. But in the
Quantum Mechanical context,
We can see the rationale: the origin of (\ref{e4}) lies in the QMBH.\\
More specifically, the temporal part of the transformation (\ref{e4}) leads
to the appearance of charge in (\ref{e5}). The space part then, as is known
leads to the Kerr-Newman metric.\\
Ever since Einstein put forward his theory of Gravitation in 1915, a problem
that has vexed physicists including Einstein himself is the incorporation
of electromagnetism into the theory of gravitation. Einstein himself
said it all in his Stafford Little Lectures delivered in May 1921 at
Princeton University\cite{r17}, "... if we introduce the energy tensor
of the electromagnetic field into the right hand side of (the gravitational
field equation) we obtain (the first of Maxwell's systems of equations in
tensor density form), for the special case ($\sqrt{-g}\rho \frac{dx_\nu}{ds}
=)\tau^\mu = 0,$... This inclusion of the theory of electricity in the
scheme of General Relativity has been considered arbitrary and unsatisfactory....
a theory in which the gravitational field and the electromagnetic
field do not enter as logically distinct structures would be much preferable..."\\
The very early attempt of Hermann Weyl to which we will return very briefly
failed for this reason. As Einstein put it\cite{r17}, "H. Weyl, and recently
Th.Kaluza, have put forward ingeneous ideas along with this direction;
but concerning them, I am convinced that they do not bring us nearer to
the true solution of the fundamental problem...". Fruitless attempts over
the years lead to Pauli's famous remark that one should not put together
what God had intended to be separate viz., electromagnetism and
gravitation (cf. also ref.\cite{r18}).\\
The basic problem is that General Relativity belongs to the domain of
classical physics whereas as Einstein observed in the above quotation,
electromagnetism belongs to the domain of "elementary electrically charged
particles", that is Quantum Theory, more specifically the theory of the
electron. And, as J.A. Wheeler\cite{r14} put it, "the most evident
shortcoming of the geometrodynamic model as it stands is this, that it
fails to supply any completely natural place for spin $\frac{1}{2}$
in general and for the neutrino in particular", while "it is impossible to accept any description
of elementary particles that does not have a place for spin half."
This apart it should be remembered that the space time we speak of in
general relativity is not only deterministic, but we also speak in terms
of definite points of space time. This as seen above is forbidden in Quantum Theory by
the Uncertainity Principle.  Infact four dimensional space time exists
only as a classical approximation \cite{r19,r20}.\\
On the other hand, Quantum Gravity wherein one invokes phenomena at the scale of the
Planck length, that is $10^{-33}cms$, has not been satisfactorily concluded.
The above Planck length considerations as is well known and briefly alluded
to earlier, lead
to what have been sometimes termed as, "maximons" particles which are
essentially Schwarzchild black holes with the Planck length radius, and with
mass $\sim 10^{-5}gm$. Such particles are too massive to be termed
"elementary", are transient, with life times $\sim 10^{-42}sec$ and moreover they do not exhibit the all important spin.\\
However the following question is to be clarified: How can an electron
described by the Quautum Mechanical Dirac spinor be identified with the
geometrodynamic Kerr-Newman black hole characterized by curved space time?
The answer is as follows:\\
As we approach the Compton wavelength the "negative energy" two rowed
component $\chi$ of the full four rowed Dirac spinor $(\theta_\chi)$, where
$\theta$ denotes the positive energy two spinor\cite{r10}, begins to
dominate. Further under reflections, while $\theta$ goes to $\theta, \chi$
behaves like a psuedo-spinor (cf.ref.\cite{r21}),
$$\chi \to - \chi$$
Hence the operator$\frac{\partial}{\partial x^\mu}$ acting on $\chi$, a
density of weight $N = 1,$ has the following behaviour,
$$\frac{\partial \chi}{\partial x^\mu} \to \frac{1}{\hbar} [\hbar \frac{\partial}
{\partial x^\mu} - NA^\mu] \chi$$
where,
\begin{equation}
A^\mu = \hbar \Gamma_\sigma^{\mu \sigma} = \hbar \frac{\partial}{\partial x^\mu}
log (\sqrt{|g|}) \equiv \nabla^\mu \Omega\label{e7}
\end{equation}
We can easily identify $NA^\mu$ in (\ref{e7}) with the electromagnetic
four potential. The fact that $N = 1$ explains why the charge is discrete.\\
In this formulation, electromagnetism is the result of the covariant
derivative that arises due to the Quantum Mechanical behaviour of the negative
energy components within the Compton wavelength. Thus we have endowed the
electron with curvature and have introduced the double connectivity of
spin half into a geometrodynamic formulation. This may be called a Quantum Mechanical
Black Hole (QMBH).\\
Equation (\ref{e7}) is apparently identical to Weyl's formulation for the
unification of electromagnetism and gravity. But there is a fine print.
Weyl's Christofell symbol contains two independent entities - the metric
tensor $g^{\mu v}$ \underline{and} the electromagnetic potential $\phi$\cite{r22}.
In effect there is no unification of electromagnetism and gravity, which
prompted Einstein's remark quoted earlier. In our formulation this follows from
the Quantum Mechanical pseudo spinor property.\\
However, to complete
our task we still have to show that the gravitational field and the
electromagnetic field, \underline{both} emerge from the above formulation.
We will first demonstrate the emergence of the mass and the charge.\\
We start with the metric of a relativistic "fluid" in the linearized
theory\cite{r14} (in units $G = 1, c=1, \hbar = 1$),
\begin{equation}
g_{\mu v} = \eta_{\mu v} + h_{\mu v}, h_{\mu v} = \int \frac{4T_{\mu v}
(t - |\vec x - \vec x'|,\vec x')}{|\vec x - \vec x'|} d^3 x'\label{e8}
\end{equation}
In the expression (\ref{e8}) there is no restriction on the velocity, while
the stresses $T^{jk}$ and momentum densities $T^{oj}$ can be comparable
to the energy momentum density $T^{oo}$.\\
As is well known when $|\frac{\vec x'}{r}| < < 1,$ where $r \equiv |\vec x |$,
and in a frame of reference with origin at the centre of mass at rest
with respect to the particle we have
$$
m = \int T^{oo} d^3 x$$
$$
S_k = \int \epsilon_{klm} x^l T^{mo} d^3 x$$
where $m$ is the mass and $S_k$ is the angular momentum, and
\begin{equation}
T^{\mu v} = \rho u^\mu u^v\label{e9}
\end{equation}
If we now work in the Compton wavelength region, we have\cite{r9},
\begin{equation}
|u^l| = 1\label{e10}
\end{equation}
(This is the Quantum Mechanical input) Substitution of (\ref{e9}) and
(\ref{e10}) in (\ref{e8}) gives on using the Mean Value Theorem,
$$S_k =  < x^l > \int \rho d^3 x$$
As $< x^l > \sim \frac{1}{2m}$, using (\ref{e9}), we get, $S_k \approx
\frac{1}{2}$, as required for a spin half particle.
This is reminiscent of and bears
a superficial resemblance to Dirac's membrane model \cite{r23}.\\
The gravitational potential can similarly be obtained from (\ref{e8}) and
(\ref{e9}) (cf.ref.\cite{r14}).
$$\Phi = - \frac{1}{2} (g^{oo} - \eta^{oo}) = - \frac{Gm}{r} + 0 (\frac{1}{r^3})
$$
Now using the expression for the Christoffel symbols, in (\ref{e7}) we have,
$$A_\sigma = \frac{1}{2}(\eta^{\mu v} h_{\mu v}),\quad _\sigma,$$
so that, from (\ref{e8}),
$$A_o = 2 \int \eta^{\mu v} \frac{\partial}{\partial t} [\frac{T_{\mu v}
(t -|\vec x - \vec x'|, \vec x')}{|\vec x - \vec x'|}] d^3 x'$$
Remembering that $|\vec x - \vec x'| \approx r$ for the distant region we
are considering, we have,
$$A_o \approx \frac{2}{r} \int \eta^{\mu v} [\frac{\partial}{\partial \tau}
T_{\mu v} (\tau,\vec x').\frac{d}{dt}(t - |\vec x - \vec x'|)] d^3 x'
$$
or finally
\begin{equation}
A_o \approx \frac{2}{r} \int \eta^{\mu v} \frac{d}{d\tau}T_{\mu v} d^3 x'\label{e11}
\end{equation}
where we have used the fact that in the Compton wavelength
region, $u_\nu = 1(=c)$\\
It has already been observed that QMBH can be treated as a rotating shell
distribution with radius $R \equiv \frac{1}{2m}.$ So we have,
\begin{equation}
|\frac{du_v}{dt}| = |u_v|\omega\label{e12}
\end{equation}
where $\omega$, the angular velocity is given by,
\begin{equation}
\omega = \frac{|u_v|}{R} = 2m\label{e13}
\end{equation}
Finally, on using (\ref{e12}), (\ref{e13}) in (\ref{e11}), we get, on
restoring usual units,
\begin{equation}
\frac{e'e}{r} = A_o \sim \frac{\hbar c^3}{r} \int \rho \omega d^3 x' \sim
(Gmc^3)\frac{mc^2}{r}\label{e14}
\end{equation}
where $e' = 1 esu$ corresponds to the charge $N = 1$ and $e$ is the test
charge.\\
We now test the validity of (\ref{e14}): Because of the approximations taken in deducing (\ref{e14}), a
dimensional constant $(\frac{L}{T})^5$ has to be multiplied on the left
side, which then becomes, in units, $c = G = 1,$
$$e'e. (\mbox{dimensional}\mbox{constant}) \approx 1.6 \times 10^{-111}cm^2$$
The right side is,
$$Gm^2c^5 \approx 4.5 \times 10^{-111}cm^2,$$
in broad agreement with the left side.\\
Alternatively, using the values of $G,m$ and $c$ in (\ref{e14}), we get,
$$e \sim 10^{-10} esu,$$
which is correct.\\
Yet another way of looking at (\ref{e14}) is, that we get, as $e' = 1 esu
\sim 10^{10}e$
\begin{equation}
\frac{e^2}{Gm^2} \sim 10^{40},\label{e15}
\end{equation}
that is, we have theoretically deduced a result
which is well known empirically.\\
Finally given the mass and charge we take the following short cut to the full
Kerr-Newman metric by considering the imaginary displacement (\ref{e4})
As is well known, in General Relativity this leads us inexplicably
and as if by accident\cite{r15,r16}, from the
static Kerr-Metric to the full Kerr-Newman metric. Newman himself observed,
"...one does not understand why it works. After many years of study
I have come to the conclusion that it works simply by accident".\\
However in the preceding setting of covariant derivatives in a Quantum
Mechanical context one can see the rationale for the transformation
(\ref{e4}): We take $a^o \approx \frac{\hbar}{2mc^2}$, and $a^\imath \approx
\frac{\hbar}{2mc}$. In other words we are probing the Zitterbewegung region of the
Quantum Mechanical Black Hole. This then leads to (\ref{e7}).
This again indicates how electromagnetism or the Kerr-Newman metric for
the electron emerge from the Zitterbewegung region.\\
In any case remembering that the Kerr-Newman field is a stationary field,
and to this extent we are dealing with the approximation of an isolated
electron at rest, it can be shown shown that the electromagnetic mass of the
Kerr-Newman field coincides with the mass itself, in this theory. This was
not possible in the Classical Theory of the electron\cite{r24}.\\
It is worth mentioning that spin can be considered to be the angular
momentum within the Zitterbewegung region\cite{r25}, while the electron
(and its massless version, the neutrino) have been considered to be the
primary elementary particle\cite{r26}. Hestenes too has argued that
the inertial energy could be considered as due to Zitterbewegung processes
within the Compton wavelength though the general relativistic
scheme is absent in this case. Finally the emergence of the inertial
mass from within the Compton wavelength can be demonstrated without
recourse to General Relativity\cite{r9,r27}.\\
In an early paper attempting to describe elementary particles within the
framework of the gravitational field equations specifically, the
Schwarzchild and Kerr solutions,\cite{r28}, Einstein
and Rosen wondered to what extent Quantum Mechanics had been included in
their scheme. After that, much work has been done in attempts to unify
General Relativity and Quantum Theory, notwithstanding the difficulties
mentioned in earlier.\\
However the fact remains that not only are Planck scale particles with mass
of about $10^{-5}gms$ too massive to be elementary, but as pointed out
earlier, they do not exhibit the all important spin half: they are essentially
Schwarzchild Black Holes. On the other hand spinorial behaviour is a very
Quantum Mechanical characteristic - it is fermions that represent the
material content of the universe.\\
Neutrinos are also described by the above model. In this case as pointed
out\cite{r9}, because they have vanishingly small mass, they have a very large
Compton wavelength so that at our usual spatial scales we encounter predominantly
the negative energy components with the peculiar handedness (cf.ref.\cite{r9}
for details).\\
It was shown in ref.\cite{r9} that if on the other hand
instead of considering distances $>>$
the electron Compton wavelength we consider the distances of the order of the
Compton wavelength itself (\ref{e8}) leads to a QCD type potential,
\begin{eqnarray}
4 \int \frac{T_{\mu \nu} (t,\vec x')}{|\vec x - \vec x' |} d^3 x' +
(\mbox terms \quad independent \quad of \quad \vec x),\nonumber \\
+ 2 \int \frac{d^2}{dt^2} T_{\mu \nu} (t,\vec x')\cdot |\vec x - \vec x' |
d^3 x' + 0 (| \vec x - \vec x' |^2) \propto - \frac{\propto}{r} + \beta r\label{e16}
\end{eqnarray}
The above considerations immediately lead us to consider the possibility of describing
weak interactions on the one hand and quarks on the other
in the above model.\\
We will now indicate how it is possible to do so \cite{r29,r30}.\\
Let us start with the electrostatic potential given in equations (\ref{e7})
and (\ref{e11}).\\
We will first show how the characteristic and puzzling $\frac{1}{3}$ and
$\frac{2}{3}$ charges of the quarks emerge.\\
For this we first note that the electron's spin half which is correctly
described in the above model of the Kerr-Newman Black Hole, outside the
Compton wavelength  automatically implies three spatial dimensions\cite{r14}.
This is no longer true as we
approach the Compton wavelength in which case we deal with low space dimensionality.
So for the Kerr-Newman
Fermions spatially confined to distances of the order of their Compton
wavelength or less, we consider two and one spatial
dimensionality.\\
Using now the well known fact\cite{r9} that each of the $T_{\imath j}$ in
is given by $\frac{1}{3} \epsilon, \epsilon$
being the energy density, it follows from (\ref{e11}) that the
particle would have the charge $\frac{2}{3} e$ or $\frac{1}{3}e$, as in the
case of quarks. Moreover, as noted earlier (cf.ref.\cite{r8,r9} also), because we are
at the Compton wavelength scale, we encounter predominantly the components
$\chi$ of the Dirac wave function, with opposite parity. So, as with neutrinos,
this would mean that the quarks would display helicity, which indeed is true:
As is well known, in the $V-A$ theory, the neutrinos and relativistic quarks
are lefthanded while the corresponding anti-particles are right handed (brought
out by the small Cabibo angle). This also automatically implies that these fractionally
charged particles cannot be observed individually because they are by their
very nature spatially confined. This is also expressed by the confining part
of the QCD potential (\ref{e16}). We come briefly to this aspect now.\\
Let us consider the QCD type potential (\ref{e16}). To facilitate comparison
with the standard literature\cite{r5}, we multiply the left hand expression
by $\frac{1}{m}$ (owing to the usual factor $\frac{\hbar^2}{2m})$ and also
go over to natural units $c = \hbar = 1$ momentarily. The potential then
becomes,
\begin{equation}
\frac{4}{m} \int \frac{T_{\mu v}}{r} d^3 x + 2m \int T_{\mu v} r d^3x \equiv
-\frac{\propto}{r} + \beta r\label{e17}
\end{equation}
Owing to (\ref{e9}), $\propto \sim O(\ref{e1})$ and $\beta \sim O(m^2),$
where $m$ is the mass of the quark. This is indeed the case for the QCD
potential (cf.ref.\cite{r5}). Interestingly, as a check, one can verify
that, as the Compton wavelength distance $r \sim \frac{1}{m}$ (in natural
units), the energy given by (\ref{e17}) $\sim O(m)$, as it should be.\\
Thus both the fractional quark charges (and handedness) and their masses are seen to arise from
this formulation.\\
To proceed further we consider (\ref{e11}) (still remaining in natural units):
\begin{equation}
\frac{e^2}{r} = 2Gm_e \int \eta^{\mu v}\frac{T_{\mu v}}{r}d^3x\label{e18}
\end{equation}
where at scales greater than the electron Compton wavelength, $m_e$ is the
electron mass. At the scale of quarks we have the fractional charge and
$e^2$ goes over to $\frac{e^2}{10} \approx \frac{1}{1370} \sim
10^{-3}$.\\
So we get from (\ref{e18})
$$\frac{10^{-3}}{r} = 2Gm_e \int \eta^{\mu v} \frac{T_{\mu v}}{r} d^3 x$$
or,
$$\frac{\propto}{r} \sim \frac{1}{r} \approx 2G.10^3 m_e \int \eta^{\mu v}
\frac{T_{\mu v}}{r} d^3x$$
Comparison with the QCD potential and (\ref{e18}) shows that the now fractionally charged
Kerr-Newman fermion, viz the quark has a mass $\sim 10^3 m_e$, which is
correct.\\
If the scale is such that we do not go into fractional charges, we
get from (\ref{e18}), instead, the mass of the intermediary particle as
$274 m_e,$ which is the pion mass.\\
All this is ofcourse completely consistent with the physics of strong
interactions.\\
It has already been noted that in the formulation of leptons as Kerr-Newman
Black Holes, for Neutrinos, which have vanishingly small mass, if at all, so
that their Compton wavelength is infinite or very large, we encounter
predominantly the negative energy components of the Dirac spinor. It was
shown in, for example, ref.\cite{r9}, that this explains their characteristic
helicity and two component character. One should expect that from the
above considerations, we should be able to explain weak interactions also.\\
Even in the early days leading to the electro-weak theory,\cite{r31}, it
was realized, that with the weak coupling constant set equal to the electromagnetic
coupling constant and with a massive intermediary particle $m_w \sim 100 m_p,$
where $m_p$ is the proton mass, we get the Fermi local weak coupling constant,\\
\begin{equation}
G_w = g^2/m^2_w \approx \frac{10^{-5}}{m^2_p}gm^{-2}\label{e19}
\end{equation}
This is also the content of our argument: We propose to show now that (\ref{e19})
is consistent with (\ref{e15}).
However, it should be borne in mind that now (\ref{e15}) is not an adhoc
experimental result, but rather follows from our model as seen earlier.\\
Our starting point is equation (\ref{e15}), which we rewrite, as
$$\frac{e^2\times 10^{19} \times 10^8}{10^4 m^2_p} \approx \frac{10^{40}\times
10^{19}}{10^4}\approx 10^{55},$$
remembering that $G \sim 10^{-8}$ and $e^2 \sim 10^{-19}$. From here we get,
$$\frac{g^2}{m_w^2} \approx 10^{43} gm^{-2}$$
with $g^2 \approx 10^{-1}$ and $m_w \approx 100 m_p$ which is
consistent with (\ref{e19}).\\
\section{Discussion}
The model of leptons as Kerr-Newman Black Holes, as discussed above
leads to a cosmology\cite{r8,r32} in which the universe continues to
expand and accelerate with ever decreasing density. Interestingly this
has been confirmed by several independent recent observations\cite{r33}.\\
On the other hand the model also predicts that near Compton wavelength scales
electrons would display a neutrino type bosonization or low dimensionality\cite{r34}.
This type of behaviour has been noticed, already at nano scales
in recent experiments with nano tubes \cite{r35,r36}.\\
What we have shown above is that the Kerr-Newman Black Hole description
of the electron leads to a unification of gravitation and electromagnetism
as expressed by e.g. (\ref{e15}). It also gives a clue to the
peculiar fractional charges and also masses of the quarks and the QCD interaction.
Finally the model explains the handedness
of the neutrino and gives a clue to the origin of weak interactions.\\
So at the heart of the matter is the description of the electron as a
Kerr-Newman Black Hole, valid at length scales greater than the Compton
wavelength. The other phenomena appear from here at different length (or
energy) scales.

\end{document}